\documentclass[preprint,12pt]{elsarticle}

 \usepackage{graphics}
 \usepackage{graphicx}
\usepackage{epsfig}
\usepackage{epstopdf}
\usepackage[cmex10]{amsmath}
\usepackage{amssymb}
\usepackage{amsthm}
\usepackage{rotating}

\journal{ICCN-2013/ICDMW-2013/ICISP-2013}

\begin{document}

\begin{frontmatter}

\title{Performance Evaluation of ECC in Single and Multi Processor Architectures on FPGA Based Embedded System}

\author{Sruti Agarwal\corref{cor1}\fnref{fn1}}

\author{Sangeet Saha\fnref{fn2}}
 \author{Rourab Paul\fnref{fn3}}
\author{Amlan Chakrabarti\fnref{fn4}}  
                                                     
\cortext[cor1]{Corresponding author}                             
\fntext[fn1]{Institute of RadioPhysics and Electronics, University of Calcutta, Kolkata.}
\fntext[fn2]{CSE, University of Calcutta, Kolkata.}
\fntext[fn3]{A.K.Choudhury School of I.T., University of Calcutta, Kolkata.}
\fntext[fn4]{Senior Memeber IEEE, A.K.Choudhury School of I.T., University of Calcutta, Kolkata.}

\address{}

%

\begin{abstract}

Cryptographic algorithms are computationally costly and the challenge is more if we need to execute them in resource constrained embedded systems. Field Programmable Gate Arrays (FPGAs) having programmable logic devices and processing cores, have proven to be highly feasible implementation platforms for embedded systems providing lesser design time and reconfigurability. Design parameters like throughput, resource utilization and power requirements are the key issues. The popular Elliptic Curve Cryptography (ECC), which is superior over other public-key crypto-systems like RSA in many ways, such as providing greater security for a smaller key size, is chosen in this work and the possibilities of its implementation in FPGA based embedded systems for both single and dual processor core architectures involving task parallelization have been explored. This exploration, which is first of its kind considering the other existing works, is a needed activity for evaluating the best possible architectural environment for ECC implementation on FPGA (Virtex4 XC4VFX12, FF668, -10) based embedded platform. 

\end{abstract}

\begin{keyword}
FPGA, ECC, single-core, dual-core, MicroBlaze, shared memory
\end{keyword}

\end{frontmatter}
\section{Introduction}
\label{section:introduction}
ECC is an approach to public-key cryptography based on the algebraic structure of elliptic curves over finite fields. The use of elliptic curves in cryptography was suggested independently by Neal Koblitz\cite{koblitz}  and Victor S. Miller\cite{miller} in 1985. Elliptic curve cryptography, in essence, entails using the group of points on an elliptic curve as the underlying number system for public key cryptography. The primary reason for using elliptic curves as a basis for public key crypto-systems is that elliptic curve based crypto-systems appear to provide better security than traditional crypto-systems for a given key size\cite{acm}, thereby reducing the process overhead. One can take advantage of this fact to increase security, or (more often) to increase performance by reducing the key size while keeping the same security.

Designers are now working on designing dedicated hardware accelerator blocks along with the main processor \cite{coprocessor}\cite{hardware}\cite{multi} to increase the throughput of the design. ECC algorithm is used for secured communication in smart cards\cite{scard} and also in GSM security\cite{gsm}. So, high speed, resource constrained  environment is required. Using a dual-core instead of a dedicated co-processor enables the user to operate from the application layer without entering the subordinate layers. Also, the thread-level parallelism, used by the dual-core ensures higher throughput without increasing the much power, which is an important issue for low-power communication devices like smart cards.  Implementations can be made in different platforms namely, FPGA or ASIC or can be done using micro-controllers. FPGAs provides reconfigurability and lesser design time, while ASIC provides better throughput though the design time is large and expensive. We propose the design and implementation of Elliptic Curve Cryptography (ECC) encryption algorithm by developing suitable single core and dual core design on Xilinx Virtex 4 (ML403) device. The system is optimized in terms of execution speed. We perform a trade-off between throughput, power and resource requirements for our dual core implementation. To the best of our knowledge, dual-soft core processor based implementation of ECC in an FPGA  is not yet available in related literatures and hence it is first of its kind.

The paper is organized as follows: Section 2 details the overview of ECC, the encryption and decryption process. The design and implementation details for single and dual-core processor architectures are described in Section 3. The experimental results are summarized in Section 4. Conclusion and References are briefed in Section 5.

%

\section{Background}                                                                            
This section provides some background on elliptic curves and ECC and then the hardness of decrypting the elliptic curve ciphers is also discussed. The idea of a multi processor system, establishment of communication between the processors using shared memory and the needed data synchronization is also briefed in the later part of this section.
\subsection{Elliptic Curves}
\begin{figure}[htbp]
	\centering
		\includegraphics[scale=.4]{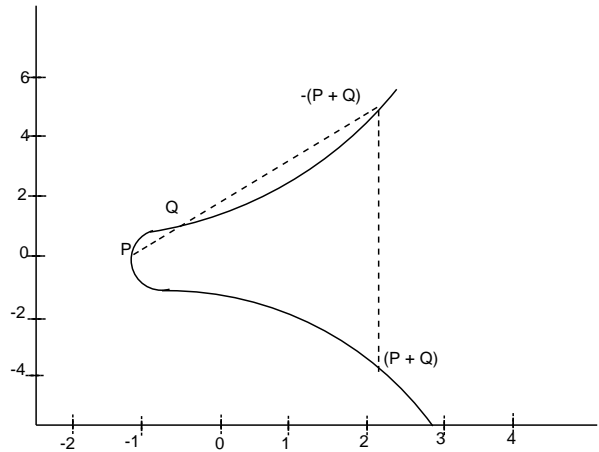}
	\caption[elliptic curve(a)]{Example of Elliptic Curve}
	\label{fig:elliptic curve(a)}
\end{figure}

Elliptic curves are described by curves, which are similar to cubic equations, used for calculating the circumference of an ellipse. In general, cubic equations for elliptic curves take the following form known as Weierstrass equation\cite{stalling}:
\begin{equation}
(y^2 + axy + by)  \;\;\text{mod p} = (x^3 + cx^2 + dx + e) \;\;\text{mod p}
\label{eqn:Weierstrass equation}
\end{equation}
where \emph{a}, \emph{b},\emph{c}, \emph{d}, \emph{e}, \emph{p} are real numbers and \emph{x} and \emph{y} take on any values in the real numbers. For our purpose, it is sufficient to limit ourselves to equation of the form given in Equation\ref{eqn:reduced Weierstrass equation} for appropriate curve parameters of ECC.
\begin{equation}
y^2 \;\;\text{mod p} = (x^3 + ax + b)\;\; \text{mod p}
\label{eqn:reduced Weierstrass equation}
\end{equation}
 Figure \ref{fig:elliptic curve(a)} shows an example of elliptic curve.
 
\subsection{Elliptic Curve Cryptography}

ECC can be used to encrypt a plain text message \emph{M} into cipher text for secured communication. Firstly, the plain text message \emph{M} is converted into a set of finite points  P$_{M}$(\emph{x}, \emph{y}), which lie in the elliptic curve E$_{p}$(\emph{a}, \emph{b}). A generator point, \emph{G} $\in$ E$_{p}$(\emph{a}, \emph{b}), is chosen next such that the smallest value of \emph{n} for which \emph{n}\emph{G} = \emph{O} is a very large prime number. The elliptic curve E$_{p}$(\emph{a}, \emph{b}) and the generator point \emph{G} are then made public.

\ \ \ \ Let there be two parties A and B who wish to communicate using ECC.  Each user selects a private key, user A's private key is n$_{A}$, while n$_{B}$ is the private key for user B such that n$_{A}$, n$_{B}$  \textless \emph{n}. Then they compute their public keys. The public key  of A is P$_{A}$ =  n$_{A}$\emph{G}, while for B the public key  is P$_{B}$ =  n$_{B}$\emph{G}. To send the secret message to B, A encrypts the message point P$_{M}$  by choosing a random number \emph{k} and computes the cipher text points  C$_{M}$ using B's public key. The cipher text is given by:
\begin{equation}
C_{M} = [(kG), (P_{M} + kP_{B})] 
\label{ciphertext}
\end{equation} 
On receiving the cipher text (pair of points) \emph{C$_{M}$}, B multiplies the first pair of points (\emph{k}\emph{G}) with his private key \emph{n$_{B}$}, and then adds the result to the second point of the cipher text (\emph{P$_{M}$ + kP$_{B}$}) i.e.,
\begin{equation}
(P_{M} + kP_{B}) - [n_{B}(kG)] = \;\;(P_{M} + kn_{B}G) - [n_{B}(kG)] = \;\;P_{M}
\label{palintext}
\end{equation} 
Plain text message point P$_{M}$, corresponds to the message \emph{M}. It is to be observed here that only B can remove n$_{B}$(\emph{k}\emph{G}) from the second part of the cipher text. No, third party or intruder can know the message except B. Thus, ECC is very secured and can be relied for confidential communication.\\
Breaking of ECC is a ``hard problem'', which requires computing of discreet logarithm \cite{stalling}. 
\subsection{Multi processor system} 
\label{mps}
A multi processor system consists of two or more processors working concurrently and capable of communicating with each other. Such a design tends to double the throughput, with two processors running independently, but with an extra cost of resource and power. On the other hand, a multi core system is one in which more than one processor is build on the same die. In FPGA we have on-chip soft processor cores, which has been utilized in our design to  multi core design. There are some basic conditions required for the execution of a design in a multi core system. The primary being concurrency in the design i.e., no data dependency must be present in processes that run in different processors and also the two processors must have some handshaking for synchronization of data.  Multi core processors often use a shared memory system or a Mailbox system as a interprocessor communication mechanism, that operates very quickly\cite{multicore}.
 \subsection{ Shared Memory }
\label{shared}
 In a multi core environment, mailbox and shared memory \cite{mailbox} are the two mechanisms, provided by Xilinx. Out of these two mechanisms, shared memory is the most common and most intuitive way of passing information between processing subsystems and we have used it in our design.
\par A shared memory system has the following 
properties\cite{mailbox}:\\
\textbullet Any processor can refer any location in the  shared memory directly by some system call.\\
\textbullet Location of data in memory is transparent to the programmer. Data could be distributed across multiple processors, with the help of some proper API, data can be handled at program level.\\
\textbullet A synchronization is a must for accessing the shared memory segment by some
hardware/software protocol between the two processors.\\

Shared memory is typically the fastest asynchronous mode of communication,
especially when the information to be shared is large. Shared memory
gives another approach of "in-place" message processing schemes. Shared
memory can be built out of on-chip local memory like BRAM or on external memory like DDR SDRAM.
\subsection{Synchronization}
\label{sync}
The region in which the shared data is stored is known as a Critical Region
in operating system terminology. Unless there is some sort of well-defined non-conflicting way in
which each processor accesses the shared data, the multi core system cannot work properly. A synchronization protocol or construct
is usually required to serialize accesses to the shared resource.
\par The XPS Mutex synchronization primitive is used in this work, which is provided by Xilinx as a separate IP-core\cite{dual}.
When using Shared memory as a method of data communication, the pseudo code should look like this to ensure proper synchronization,\\
{\em /* shared tasks */}\\
{\em XMutex\_Lock ();}\\
{\em /* Critical Region - Perform shared memory access */}\\
{\em XMutex\_Unlock ();}\\

By calling the {\em XMutex\_Lock()} it must be ensured that one processor is accessing the critical region and other processor should not be allowed to access the same until {\em XMutex\_LocK ()} is called by that processor.

\vspace{-10pt}
\section{Design and Implementation}
This section highlights the key components of our proposed design. At first, we describe the processor that is used and then we brief the design innovations made to enhance the throughput.
\subsection{MicroBlaze}
The MicroBlaze embedded processor soft core is a 32-bit Reduced Instruction Set Computer (RISC) optimized for implementation in FPGAs\cite{mbl}. As a soft-core processor, MicroBlaze is implemented entirely in the general-purpose memory and logic fabric of Xilinx FPGAs. The Embedded Development Kit (EDK) platform from Xilinx has been used to build a complete processor system on FPGA.

\subsection{ ECC in Single MicroBlaze}
\begin{figure}[htbp]
\vspace{-40pt}
\centering
\includegraphics[height=6.0cm,width=9.0cm]{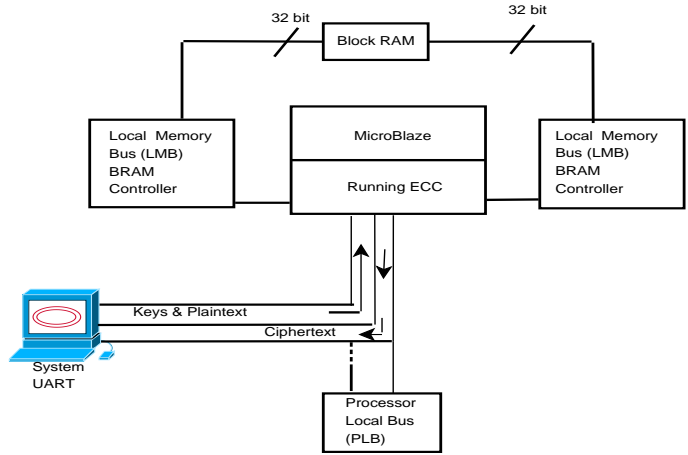}
\caption{Single MicroBlaze Architecture}
\label{single}
\vspace{-14pt}
\end{figure}

In a single processor architecture, the FPGA receives the input via the RS-232 port through UART and then the input plaintext message is encrypted using ECC algorithm, which is running in the MicroBlaze processor of the the FPGA. After encryption, the cipher text is send back to the Host PC and is seen in the Hyper Terminal using serial communication between the board and the PC. Figure \ref{single} shows the flow of steps in the design. Processor working at 100MHz clock frequency is used to encrypt 8-bits of message using ECC.  Scalar multiplication constitutes the main operation in ECC. It is seen that the processor takes 19.01 msec to perform the total encryption. The resource utilized i.e., the number of LUTs and slices required by the design as well as the power requirements are summarized in the table \ref{table:Resource and power calculation}.

\subsection{ ECC in Dual MicroBlaze}
\begin{figure}[htbp]
\centering
\includegraphics[height=6.0cm]{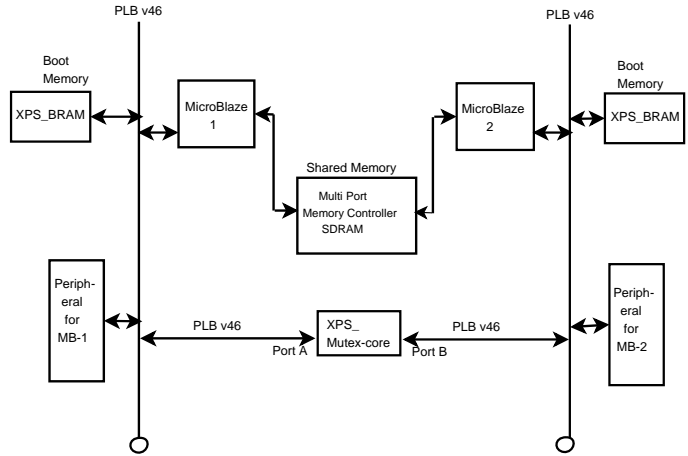}
\caption{Dual MicroBlaze Architectural flow}
\label{dual}
\vspace{-12pt}
\end{figure}

After evaluating the performance of ECC in single processor architecture by using MicroBlaze  soft-core processor, another approach is taken to implement it in Dual processor architecture by considering the fact discussed in Section ~\ref{mps} for higher throughput and efficiency. Xilinx Virtex4 ML403 is taken as a platform for design. Shared memory is used for creating a handshaking between the two processors as described in Section~\ref{shared} and the synchronization for communication is achieved using the process described in Section~\ref{sync}. 
Fig.~\ref{dual} shows the architecture of Dual MicroBlaze design,

In addition, multiple port memory controller (DDR\_SDRAM) and inter-processor communication-XPS Mutex hardware IP
\cite{dual} is also incorporated in the test bed to facilitate the memory sharing and inter processor synchronization and communication.
\par
\begin{figure}[htbp]
\vspace{-10pt}
\centering
\includegraphics[height=7.0cm,width=11.0cm]{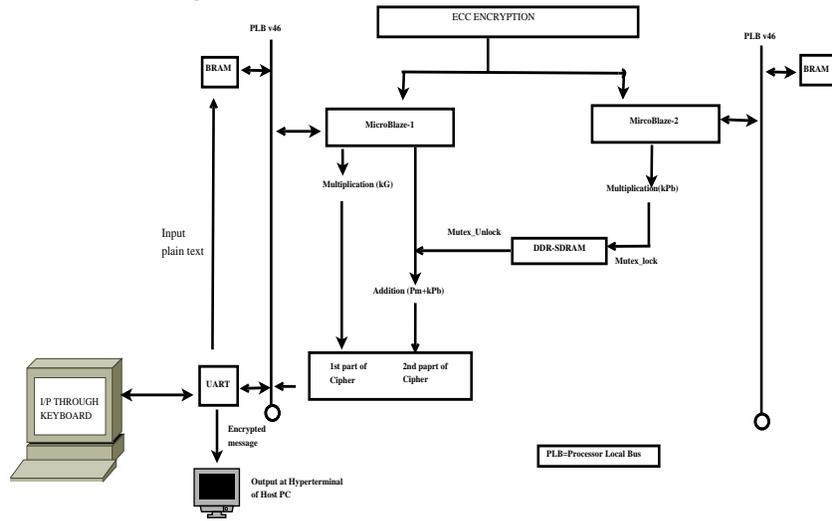}
\caption{ECC in Dual MicroBlaze}
\label{dual-mb}
\vspace{-20pt}
\end{figure}

 In the experiment, the two processors are equally engaged to execute ECC in parallel fashion and in between the consequent steps, the two processors communicate with each other via shared memory with proper synchronization. The message to be encrypted is transferred to the FPGA from a computer using RS232 serial port communication. The two multiplications are executed concurrently in two processors and the resulting data is assembled in the shared memory. The addition operation is then performed by the 1st processor and the resultant cipher text is generated. The resultant cipher text is send back to the PC using the RS232 interface and is viewed at the Hyper Terminal of the computer. Fig.~\ref{dual-mb} shows the mechanism in which the two processors communicate using the shared memory and using Mutex locking and unlocking.

\section{Result and Analysis}
It is seen that the encryption engine speeds up by 3.3 times as the proposed architecture takes only 5.72 msec to perform the encryption.The resource estimation, power required and throughput measurements for both the designs are shown in Table \ref{table:Throughput calculation} and Table \ref{table:Resource and power calculation} below. The design improves the throughput of execution, but utilizes more resources due to the dual cores. It is to be noted here that the throughput of the design can be improved further by enabling the cache memory. But due to the resource constraints of our implementation board the cache memory of the two processors could not be enabled.

\begin{table}[htbp]
\caption{For an encryption process} 
\centering 
\begin{tabular}{|c |c |c |c |c|} 
\hline\hline 
 Architecture & Clock Freq. & \# clock & Time Reqd. & Throughput \\ [0.5ex] 
  & (in MHz) & cycles reqd. & (in msec) & (bits per sec) \\[0.5ex]
\hline\hline
Single-core Microblaze & 100 & 1901317 & 19.01 & 420.83\\
Dual-core Microblaze \ & 100 & 572552 & 5.72 & 1398.60\\
\hline 
\end{tabular}
\label{table:Throughput calculation} 
\end{table}
\vspace{-20pt}
\begin{table}[htbp]
\caption{Resource and Power Estimation for the encryption process} 
\centering 
\begin{tabular}{|c| c| c| c| c| c| c|} 
\hline\hline 
 Processor & Clock Freq. & \# & \# Slice & \# 4-input & Through & Total Power\\ [0.5ex] 
  & (in MHz) & Slices &  FFs &  LUTs & put per slice & (in Watts) \\[0.5ex]
\hline\hline
Single-core & 100 & 3580 & 3750 & 4076 & 0.1175 &1.106\\
Microblaze &  &  &  &  &  &\\
Dual-core & 100 & 5313 & 6637 & 7495 & 0.2632 & 1.808\\
Microblaze &  &  &  &  &  &\\
\hline 
\end{tabular}
\label{table:Resource and power calculation} 
\end{table}

\section{Conclusion and Future Work}
This work is an exploration of ECC implementation for FPGA based embedded systems. Two specific designs have been addressed, the first one is simpler involving a single Microblaze core whereas the other one utilizes two Microblaze cores and thus enables multi-threading. The dual core based implementation gives almost a 3 times increase in the throughput but utilizes almost twice the resource and 30\% more power as compared to the single core based implementation. This clearly shows a trade off between speed, resource utilization and power requirement. In future we look forward to do a further exploration of ECC implementation for embedded applications involving FPGA based hard processor cores and ASIC based design. 

\end{document}